\documentclass[conference]{IEEEtran}

\usepackage{cite}
\usepackage{lineno}
\usepackage{amsmath,amssymb,amsfonts}
\usepackage{algorithmic}
\usepackage[graphicx]{realboxes}
\usepackage{multirow}
\usepackage{tabularx}
\usepackage{color}
\usepackage{url}
\usepackage{booktabs}
\usepackage[normalem]{ulem}
\useunder{\uline}{\ul}{}
\usepackage{afterpage}
\usepackage{amssymb}

\usepackage{textcomp}
\usepackage{xcolor}
\usepackage{cite}

\usepackage{rotating}
\usepackage{adjustbox}
\usepackage{lscape} 

\usepackage{enumitem}
\usepackage{graphicx}

\usepackage{eso-pic}

\usepackage[ruled]{algorithm2e}
\modulolinenumbers[5]

\newcolumntype{L}[1]{>{\raggedright\let\newline\\\arraybackslash\hspace{0pt}}m{#1}}  

\begin{document}
\AddToShipoutPictureBG*{%
  \AtPageUpperLeft{%
    \setlength\unitlength{1in}%
    \hspace*{\dimexpr0.5\paperwidth\relax}
    \makebox(0,-0.35)[c]{\large Author's accepted pre-print - IN PRESS}
    \makebox(0,-0.8)[c]{\large IEEE  International Conference on Cyber Security and Protection of Digital Services (Cyber Security 2020)}%
}}

\title{Towards Identifying Human Actions, Intent, and Severity of APT Attacks Applying Deception Techniques - An Experiment}

\author{
\IEEEauthorblockN{Joel Chacon}
\IEEEauthorblockA{\textit{Eigen Ltd} \\
Surrey, England \\
joel.chacon@eigen.co}
\and
\IEEEauthorblockN{Sean McKeown}
\IEEEauthorblockA{\textit{School of Computing} \\
\textit{Edinburgh Napier University}\\
Edinburgh, Scotland \\
s.mckeown@napier.ac.uk}
\and
\IEEEauthorblockN{Richard Macfarlane}
\IEEEauthorblockA{\textit{School of Computing} \\
\textit{Edinburgh Napier University}\\
Edinburgh, Scotland \\
r.macfarlane@napier.ac.uk}
}



\maketitle
\pagestyle{plain}


\begin{abstract}


Attacks by Advanced Persistent Threats (APTs) have been shown to be difficult to detect using traditional signature- and anomaly-based intrusion detection approaches. Deception techniques such as decoy objects, often called honey items, may be deployed for intrusion detection and attack analysis, providing an alternative to detect APT behaviours.
This work explores the use of honey items to classify intrusion interactions, differentiating automated attacks from those which need some human reasoning and interaction towards APT detection. Multiple decoy items are deployed on honeypots in a virtual honey network, some as breadcrumbs to detect indications of a structured manual attack. Monitoring functionality was created around Elastic Stack with a Kibana dashboard created to display interactions with various honey items. APT type manual intrusions are simulated by an experienced pentesting practitioner carrying out simulated attacks. Interactions with honey items are evaluated in order to determine their suitability for discriminating between automated tools and direct human intervention.
The results show that it is possible to differentiate automatic attacks from manual structured attacks; from the nature of the interactions with the honey items. The use of honey items found in the honeypot, such as in later parts of a structured attack, have been shown to be successful in classification of manual attacks, as well as towards providing an indication of severity of the attacks

\end{abstract}


\section{Introduction}
\label{sec:intro}

Ahmad et al.~\cite{Ahmad2019} have recently defined APTs as “An entity that engages in a malicious, organized, and highly sophisticated long-term or reiterated network intrusion and exploitation operation to obtain information from a target organization, sabotage its operations, or both”. APTs 
conduct 
stealthy operations on specifically selected target organisations and their level of sophistication includes the exploitation and even harvesting zero-day vulnerabilities. Intellectual property or productivity losses after APT attacks can have devastating consequences for both public and private organisations.

Because of these characteristics, APTs remain a major challenge for security professionals. There is evidence in literature that both signature-based and anomaly-based intrusion detection techniques may be ineffective against APTs due to either a high rate of false positives, false negatives or both ~\cite{Fraunholz2018a,RamaKrishna2019}. Regular 
server patching will also not  defend  against  
an attack employing zero-day vulnerabilities; and persistent spear-phishing 
means that it only takes one click for the attacker to set foothold inside the network of an organisation and deploy their custom-coded 
back-door tools \cite{TrendMicro2012}.

Therefore, there is still a need for new and effective APT attack detection paradigms and techniques applicable in practice, over a wide range of organisations and industries. In this context, the use of deception techniques has gained recent  popularity~\cite{Musich2019, Dunn2019} and also recently explored by the MITRE Corporation~\cite{Schuh2019}, and so has  been  
selected  as  the technique to be explored and investigated in this work.

The original deception device: the honeypot, was developed over 30 years ago with the purpose of attracting attackers to reveal their intentions, strategies and toolset. Most recently the use of other honey items, known more generally as "deception techniques" operate on the basis that any interaction with these resources is malicious in nature and can be detected by traditional monitoring systems.

Deception has developed over the years to include 
complex constructs such as honeynets and simplified elements known as honeytokens. The latter is a very general concept and is normally defined as a honeypot that is not a computer, but rather a sort of digital entity that has no legitimate purpose~\cite{Spitzner2003a}. 
As the role of deception techniques in cyber security increases, so will the need for an ethical framework. Early research suggests that defensive deception, such as proposed in this work, is generally ethical \cite{Rowe2008}.

This paper seeks to experimentally verify that it is possible to 
detect and classify intrusions according to the level of human reasoning and interaction required to execute or set them up, making use of honeytokens within a self contained system. This is in contrast to prior work which does not experimentally evaluate the use of honeytokens for intrusion detection\cite{Spitzner2003a,Almeshekah2015,Zhao2013,Virvilis2014}. Insights gained from this work are also useful when reasoning whether actions are carried out by a human actor, as honeytokens are designed to require human reasoning, while prior work  makes use of statistical heuristics to identify automated tools~\cite{Kemppainen2018,Udhani2019}.

The objective is achieved by setting up an experimental honeypot system in a real private cloud environment to test its ability to detect intrusions and classify them as stated above.

\section{Background}
\label{sec:background}

A big challenge in preventing APT attacks lies in the effectiveness of social engineering, targeting the human element, which is notoriously difficult to mitigate technically. APTs are also inherently stealthy in their operation and constantly adapt to countermeasures developed to uncover them, 
as has been evidenced by efforts to catalogue and classify their techniques, tactics and procedures (TTPs)~\cite{MITRE2020,Lemay2018}. 

But before an APT launches a social engineering attack, they must collect as much information about the target as possible, and this presents an opportunity to detect their activity, as has been demonstrated in prior work~\cite{Virvilis2014,Paradise2017}. 

However, once the initial attack vector has been successful by means of the exploitation of a vulnerability, the attack has successfully compromised a host inside the company intranet. 
At this stage it can be said that the attacker has gained control of a host “inside the defence perimeter”, and the next stage of infection can be initiated, such as privilege escalation and lateral movements. These steps, even though aided by hacking tools, are human driven as they requires analysis of the findings and decision of where to go next. In this sense, an APT attack may share characteristics with insider threats; because the actions taken after the initial foothold are driven by human actors from compromised user accounts~\cite{Virvilis2014}.

There are two distinct traditional intrusion detection approaches:
signature-based Intrusion Detection Systems (IDS), and anomaly-based IDS: Signature-based IDS are designed to detect specific pre-defined patterns of malicious behaviour and are therefore ineffective against APTs due to the innovative, polymorphic and stealthy nature of these types of attacks. This fact is well known in the research community and therefore the focus has shifted towards anomaly-based detection techniques~\cite{Virvilis2014}, which focus on detecting intrusions by attempting to recognise patterns of unusual behaviour as reflected in host logs and network traces, generally by using machine learning or hybrid techniques combining deterministic, signature-based and artificial intelligence approaches.

However, some authors~\cite{Virvilis2014,Sommer2010} claim that anomaly-based detection systems are flawed at the root of their assumption to detect APT actors: while the machine learning paradigm assumes training based on known malicious behaviour patterns, it cannot be expected to be effective against new or unknown attack patterns. It has also been reported that anomaly-based detection systems generate a large amount of log data that needs to be examined by human analysts~\cite{Almeshekah2015}, which includes a large number of false positives. The investigation of ways to reduce the number of false positives continues to be the subject of recent research by Krishna et al.~\cite{RamaKrishna2019}. Likewise, "normal" behaviour data-sets for training are difficult to obtain~\cite{Sommer2010}. Nicho, Oluwasegun and Kamoun~\cite{Nicho2018} also concur, pointing out that APTs exploit unknown attack patterns or have carefully studied how to evade well-known security mechanisms. 

Due to the APTs level of sophistication designed to avoid detection by conventional means, researchers widely agree that these type of attacks are virtually impossible to prevent~\cite{Virvilis2014}, that an effective defence cannot rely on prevention alone~\cite{Piper2013}, and that the base assumption should be that the malicious actor has already breached the perimeter~\cite{Urias2016}. Industry analyst Gartner predicted that by 2020, prevention-centred strategies will be completely ineffective against targeted attacks~\cite{Brewer2014}, and this is certainly the consensus of the research community.

A shift in emphasis from prevention to response~\cite{Baskerville2014} necessarily requires an effective detection mechanism. A key element in countering these types of actors need to be in the early detection of their activity so that countermeasures and escalation prevention actions can be initiated as early as possible in order to limit damages.

FireEye~\cite{Fireeye2019} reported that, on average, adversaries inside a North American network remain undetected for 56 days, a figure which is even greater in other regions.
CrowdStrike~\cite{Crowdstrike2020} found that, on average in 2019, an adversary takes 9 hours to move laterally and compromise other hosts within an organisation; all of which emphasises the need for fast and early detection of breaches.

Two key advantage of deception techniques is that their false-positive rate can be significantly lower compared with other intrusion detection techniques, as any interaction is by definition not normal; and at the same time have a higher probability of reducing the false-negative rate, as any new, unknown 
pattern of attack is likely to be successful against deployed honeypots~\cite{Spitzner2003a} and therefore effectively detected~\cite{Fraunholz2018}. 

While honeypots are essentially computer hosts with a CPU at its core; honeytokens are not and take many forms. Files with decoy content can be honeytokens if they are planted throughout the network's resources and given interesting names, such as “master-passwords”, and monitored for access, as suggested by Kim and Spafford~\cite{Kim1994}.
Their work is the earliest suggestion of the possible usefulness of a honeytoken even before the term was coined.

False records can be honeytokens; such as fake user or username entries in databases: If the record is used to attempt to access additional information, it would signal an unauthorised breach~\cite{Spitzner2003a}. As Spitzner~\cite{Spitzner2003b} notes, the concept of a honeytoken can be extended beyond database entries: file, Web or email servers can all have embedded honeytokens such as unique tags detectable by traditional signature-based IDS system. 

Applied to Web Servers, Fraunholz and Schotten~\cite{Fraunholz2018} identified six deception techniques: Fake Web server banners; false entries in the robot.txt file; fake error code responses; honey or decoy files, with information leading to intrusion detection or to distract or delay any attacker, respectively; introducing adaptive delays to IP addresses making repetitive requests; false comments within the HMTL code: “honey-comments” containing any misleading information such as fake credentials and fake links. These researchers applied several of these techniques to existing production Web servers, to mislead attackers and increase the security of Web servers. Several of these techniques had already been proposed by Virvilis~\cite{Virvilis2014}.

The term "Breadcrumbs" has recently appeared in commercial literature~\cite{Bushby2019,Musich2019}, referring to a type of honeytoken planted in real production systems in order to lure attackers towards honeypots and other decoys.

Various authors claim that the use of deception techniques and honeytokens in particular can be effective against cyber-attacks, however, none of these authors claims to have demonstrated this in practice: Spitzner~\cite{Spitzner2003a} “feels” that honeytokens can be effective against actors seeking information, such as intelligence seeking or industrial espionage. Almeshekah \& Spafford~\cite{Almeshekah2015} conclude that deception techniques have been shown to be effective in many contexts because they take advantage of human biases making honey-x devices seem plausible, but do not demonstrate this experimentally. Zhao \& Mannan~\cite{Zhao2013} claim that their fake authentication and session handler \textit{Uvauth} can “effectively deceive an attacker assuming fake sessions can be efficiently generated”, but do not report results of any experimental trials and in their conclusions admit that the system has not been fully evaluated. Finally, Virvilis~\cite{Virvilis2014} , who proposes a number of simple and attractive honeytoken techniques directed at detecting APT and insider-attacks does not claim to have tested them for effectiveness.

On the other hand, 
other researchers working with low or medium interaction honeypots exposing an SSH service to the internet have explored the topic of segregating human vs bot (i.e. automated) interactions. It is well known that hosts exposed to the internet will be attacked by opportunistic actors dedicated at harvesting compute boxes in order to launch other attacks; and so these researchers logged and analysed breach attempts in their thousands. Using medium interaction honeypots that enabled fake login sessions, Kemppainen and Kovanen~\cite{Kemppainen2018} identified bots through measurable host logged features: command timestamp consistency and delay between commands (almost always 4 seconds), command execution despite errors in the previous command, and detecting the same command executed over separate sessions. Unsurprisingly, they found that only 37 interactions were executed by humans (2.3\% of the total). 

Udnahi et al.~\cite{Udhani2019} used a low interaction honeypot (only to capture login attempts), simulating a server farm of 65,000 hosts and analysed over 5 million connection requests. They characterised human actions based heuristics determined from an observation of the data: the number of requests per minute ($\leqslant 10$ requests/min), the number of targets attempted ($\leqslant 2$ per day), rate of characters in each request ($\leqslant 3$ per sec) and the speed at which the password guess is typed ($\leqslant 3$ characters/sec). However, this methodology did not consider the initiation
of a brute-force attack with particular username and password dictionaries which may be derived from earlier Open-Source Intelligence (OSINT) or reconnaissance activities on the target and would be classed as actions by a bot.

\section{Methodology}
\label{sec:methodology}

Given the lack of experimental data in the identification of human interactions in the context of an insider threat or an APT actor with a foothold inside the organisation, the research questions to be investigated are:

\textbf{RQ1:} Can the influence of human reasoning and interactions be deduced with the use of honeytokens in honeypots located within the security perimeter?

\textbf{RQ2:} Can such an arrangement of honeytokens be used to determine the level of intent or severity of the attack? 

To answer these questions, the experimental method focused on the deployment and evaluation of various honeytoken techniques in a high interaction Linux-based honeypot deployed in a private cloud environment. The honeypot will be connected and communicated to via a purposely deployed virtual network segment which also has no legitimate purpose, and is an integral part of the deception (ie a honey sub-net).

The principle of all honeytokens for intrusion detection is that any interaction or actions resulting from the observation of honeytokens represent an intrusion attempt; therefore, the experiment also required the setup of a network traffic, honeypot and honeytoken monitoring infrastructure. 
The environment and experimental design followed the steps below:
\begin{enumerate}
    \item 	Selection of a set of honeytoken techniques, suitable to address the research questions.
    \item 	Selection of a honeypot where the selected honeytoken techniques can be deployed, so that human interactions, intent and severity of the intrusion attempt can be determined.
    \item Deployment of the experimental honeypot and honeytoken arrangement in a company virtual private infrastructure within a virtual sub-net.
    \item Design and deployment of an intrusion detection harness to monitor honeypot logs and network traffic indicative of interaction with the deployed honeypots and honeytokens.
    \item  Setup of a VPN with direct access to the virtual sub-net where the honeypot is located.
    \item The design of the experiment to simulate and insider or APT attack. This was done by means of a penetration testing (pentest) engagement via the VPN.
\end{enumerate}

The experimental design includes inviting  pentesters and ask they behave as APT attackers would in the second stage of the attack; after the initial foothold within the company firewall. The hypothesis is that they would find and try to use the honeytokens as means of breaching the security of the honeynet systems thereby triggering the detection of intrusions, and human interaction in particular. 
The authors recognise that the experimental design may differ in many aspects from the behaviour of a real APT actor, e.g. the pentester is directed at a particular sub-net and given a specific pentest brief, while real attackers can decide where what and where to probe; however, it is believed that the human behaviour of an attacker upon observing the honey items may be similar to that of a pentester, which is at the core of this research.

Several honeytoken techniques were selected from the literature to help evaluate the interactions and for comparison to previous work:

\begin{enumerate}
\item Use of ``disallowed'' entries with the robot.txt file ~\cite{Virvilis2014,Fraunholz2018}
\item An invisible link (e.g. white font over white background) in a webpage~\cite{Virvilis2014}, and
\item Fake credentials added as comments within the HTML source. In the case of the fake credentials~\cite{Virvilis2014,Fraunholz2018} . This honeytoken had an additional original added feature: the username planted was an incomplete email; i.e lacking the full domain name.
\end{enumerate}

The selection of these honeytoken techniques is justified based on their potential to detect human interactions and be self-contained within the honeypot. Indeed, honeytoken 1) and 3) may also be considered breadcrumbs, because their discovery may invite a human actor to investigate further into the deception: explore the disallowed folder or try the fake credentials as a login. 
The selection of technique 3) also aligns with MITRE's technique T1081 "Credentials in Files" \cite{MITRE2020}.

The monitoring function outputs and honeytoken interactions are categorised into three types of intrusion indicator, depending on the severity of the intrusion detected:

\begin{itemize}
    \item \textbf{Low priority}, if there is any level of network detected interaction with the honeypot or the use of automatic reconnaissance tools with minimal human reasoning. These are indications of attacker reconnaissance activity or the start of an attack on the honeypot.
    \item \textbf{Medium priority}, if the results of the reconnaissance were used to take the intrusion further and attempt a breach, with some possible human actions and if they indicate sustained attack activity using automated tools.
    \item \textbf{High priority}, if honeytokens were discovered by a human actor and further reasoning was applied to expand the efforts leading to a breach, for example if the information found in one honeytoken (a breadcrumb) was used in a subsequent step, indicating a high level of intent.
\end{itemize}

Sub-classifications of these indicators are also possible to indicate degrees of severity of each type of attack detected.

If intrusions are detected, which would have to have had human associated level of prior reasoning, which were not possible or unlikely to be executed by a bot, would point towards fulfilling RQ1. If different intrusion level indicators (low, medium and high) were detected in sequence from low to high, that would indicate some success in terms of RQ2, and a high level of intent in breaching the system.

Since all these honeytokens are part of a webserver, a fake client user login application was implemented using PHP code on the honeypot, which never leads to a session and simulates variable response times to the user.

\begin{figure}[h]
    \centering
    \includegraphics{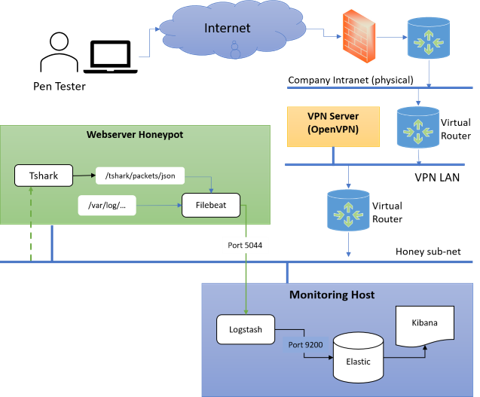}
    \caption{Experiment Set Up and Network Diagram}
    \label{fig:network}
\end{figure}

Figure~\ref{fig:network} represents the experimental set-up and monitoring infrastructure. The Elastic Stack, by Elastic, 
composed of Elasticsearch, Logsatsh and Kibana was used for the monitoring of the honeypot logs and network traffic. Filebeat, an open source log file shipper was used to monitor various system logs and TShark was used to monitor network traffic in and out of the honeypot. TShark can directly output a JSON file format directly compatible with Logstash~\cite{Wurm2017}.

Filebeat was installed in the honeypot webserver to monitor webserver (Nginx) and PHP logs, which were then ingested by Logstash on the monitoring server. The Nginx logs record access to different served pages as standard. Within the webpage PHP code, any string message can be programmaticaly directed to the PHP log. This feature was used to create log messages when the index.php page was accessed and also the user email and password entered in the login page. Filebeat was also used to monitor the packet capture JSON log file created by TShark.

The following filter (see below) was added to the TShark command line to output ICMP and TCP SYN packets to or from the honeypot, as a means to detect incoming network activity (attack) or outgoing traffic (honeypot compromise):

\begin{quote}
\footnotesize
\texttt{tshark -i eth0 -f "icmp || (tcp[tcpflags] \& (tcp-syn) != 0 and tcp[tcpflags] \& (tcp-ack) = 0)" -T ek > /home/tshark/packets.json}
\end{quote}
Logstash was configured to create indices from these data sources and push the result into the Elasticsearch database. Kibana was used to set up specific monitors to detect events of interest in a single dashboard.

\begin{figure*}[h]
    \centering
    \includegraphics[width=0.95\linewidth]{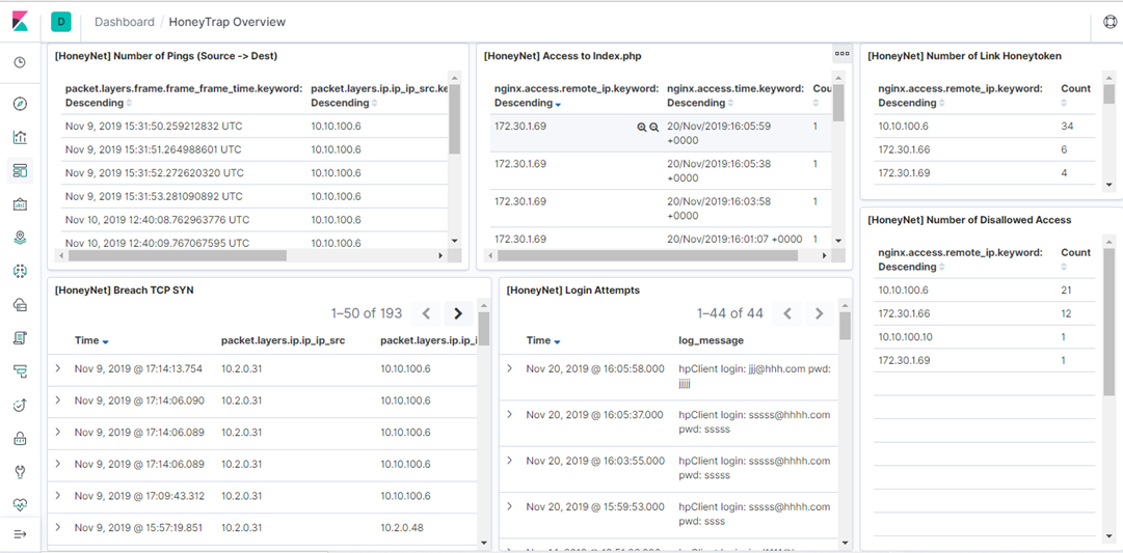}
    \caption{Kibana Dashboard}
    \label{fig:kibana}
\end{figure*}

A custom Kibana dashboard (Figure~\ref{fig:kibana}) was configured to display (from left to right, top to bottom): a) ICMP packets, b) Source IP and timestamp of Index.php access, c) No. of clicks on hidden URL, d) Timestamp, Source and Destination IP of any TCP SYN packets, e) Timestamp and credentials used in any login attempt and f) No. of accesses to disallowed folder.

Because this setup detects traffic on the honey sub-net and interactions with the honeypot, 
any event detected will have a malicious or unauthorised implication; but with varying levels of severity. 
Section~\ref{sec:analysis} (specifically Table~\ref{tab:result-table}) analyses the types of intrusions that can be detected in the system as designed, their suitability to be attacked by an automated process and their implied intent and severity.

\section{Penetration Test Process Summary}
\label{sec:findings}

An experiment 
was was carried out in January 2020, with a duration of approximately 1.5 hours. The test was conducted by an experienced industry professional pentester.

The sequence of events of the pentest was as follows, as described by the tester:

\begin{enumerate}
    \item Initial brief reconnaissance of available hosts using Nmap and including ping sweeps. Pings to the webserver (honeypot) host were captured by the monitoring harness.
    \item Version scan and OS detection using Nmap.
    \item Analysis of HTTP requests and responses. Modifying with special characters looking for SQLi and other code injections, using Burp suite.
    \item Vulnerability scan of HTTP servers, using Nikto.
    \item Directory fuzzing  of HTTP servers, using dirb.
    \item Attempt to brute force login of webserver, using Hydra.
\end{enumerate}

The pentest attack followed several stages as evidenced in the Kibana dashboard and the sequence of events in the integrated logs. The elapsed time after the test start time is indicated between brackets.

\vspace{3pt}
\textbf{Stage 1} -- Reconnaissance\newline
[+ 8 mins] During this stage, after a single ping; a flood of TCP SYN packets was detected; as shown in Figure~\ref{fig:kibana}.

\textbf{Stage 2} -- Burpsuite sweep of Web application\newline 
[+ 13 mins] The count of SYN packets reaches 173 while Burpsuite scans through the Web application pages; the main “Index.php” which presents the fake login is accessed a few times and non-authorised /admin folder is attempted 10 times. Recall the /admin folder is included in robot.txt file as disallowed; however, at this stage, it is deduced that it was attempted to be accessed by Burpsuite irrespective of its inclusion in robot.txt.

\textbf{Stage 3} -- Exploring the fake login page (Index.php)\newline 
[+ 14 mins] The count of SYN packets reaches 216 while Burpsuite continues to scan through the Web application pages; the main “Index.php” which presents the fake login is accessed 35 times and a blank login email/password combination is detected. This is clearly the continuation of the automated scan stage; given the number of SYN packets, the number of detected separate accesses to Index.php and the number of attempts to reach /admin (47 times).

\textbf{Stage 4} -- Login credentials, initial test\newline
[+ 14 mins] While the automatic scanning continues; at this stage is the first sign of human intervention with the attempted login with credentials that could be guessed from the context of the test. 

\textbf{Stage 5} -- Further automated scans\newline
[+ 17 mins] While the login attempts temporarily stop at 3; the number of SYN packets rises again, the number of accesses to Index.php rises to more than 20 every second and attempts to access /admin shoots to more than 1300. This indicates a return to automatic scanning tools, possibly dirb\footnote{A Web content scanner. See \url{https://tools.kali.org/web-applications/dirb} for more information}. This stage continues until with 783 SYN packets, 11 blank login attempts and 12,004 attempts to access the /admin folder, as shown below. As determined during the analysis of these results; this sharp increase was the result of the use of the tool dirb, fuzzing sub-folders inside /admin.

\textbf{Stage 6} -- Use of Fake Credential Honeytoken\newline
[+ 36 mins] At this stage the attacker has examined the source code of the Index.php page and identified the fake credentials added as a comment:

\noindent {\small \texttt{$<$!-- test login user: eigentest1@eigen psw: e1Ars3nal --$>$}}

And tried to use these credentials to gain access in three opportunities: The first time he makes a mistake (password “eIArs3nal” instead of “e1Ars3nal”), then he corrects this mistake; and finally completing the incorrect email reference by adding .co to the user email, as per the company’s domain (eigen.co). This stage confirms human intervention and reasoning impossible to be duplicated by a robot, at least with the current state of technology. It also demonstrates clear intent to breach the security; as the attacker followed a series of clues to try to breach the security of the Web page.

\textbf{Stage 7} -- Access to hidden link Honeytoken\newline 
[+43 mins] At this stage the hidden link was discovered and accessed. The pentester reported that this link was discovered with dirb; but dirb uses a fuzzing technique to try to guess hidden folders. A wide search across all of dirb wordlist files (command line: ``\texttt{cat * | grep testlogin}'') revealed that “testlogin” is not one of the directories that dirb could guess, unless a very customised wordlist is used.

On the other hand, when Burpsuite was directed at the root of the Web application in order to verify this result; it immediately detected the link /testlogin/index.php. Since Burpsuite was already used in Stage 2 of the pentest 
the implication is that the repeated run to discover and investigate this hidden link demonstrates a higher level of commitment and intention to attack.

\textbf{Stage 8} -- Login Brute Force\newline
[+ 66 mins] In this stage a login brute force attack was attempted, pre-configured with variations on the original login honeytoken; and common simple passwords; e.g. “123456”; and usernames such as admin@eigen.co, administrator@eigen.co and ROOT@eigen.co. An earlier login attempt with email ‘@gmail indicates a type of SQL injection.

[+ 94 mins] the credentials “eigentest1@eigen” with password “Liverpool” (another football team) was attempted; again, a variation on the planted honeytoken. By +97 mins a total of 1412 email/password combinations had been tested; many focusing on the username of “eigentest1@eigen” planted in the honeytoken; with a large number of different passwords; which indicates a premeditated configuration of the brute force attack using the fake credential username discovered prior.

[+ 98 mins] the Test concluded.

\section{Analysis}
\label{sec:analysis}

\begin{table*}[h]
\begin{center}
\renewcommand*{\arraystretch}{0.4}
\begin{tabular}{@{}|L{4cm}|L{5.5cm}|L{1.8cm}|L{1.5cm}|L{4cm}|@{}}
\toprule
\textbf{Event detected}                                                                                           & \textbf{Implication}                                                                                                                                                      & \textbf{Severity Level (Design)}  & \textbf{Auto/ Human} &
\textbf{Actual Result and Severity}                                                                                                                \\ \midrule
ICMP pings or incoming SYN traffic on Honeypot 1                                                                  & Initial stages of reconnaissance, opportunistic attacker or involuntary action by insider.                                                                                                               & Very Low  & Auto & Verified Correct.   Very Low.       (packet capture)                                                                                               \\ \midrule
Access to Index.php of the Webserver                                                                              & Initial stages of webserver reconnaissance.                                                                                                                               & Low     & Auto                          & Verified Correct.   Low.       (Nginx log)                                                                                                         \\ \midrule
Click on hidden url                                                                                               & Web application crawling                                                                                                                                                  & Medium-Low       & Auto or Human                & Verified Correct.       The link was most probably discovered by Burpsuite;   towards the end of the pentest.       (hidden /testlogin/index.php) \\ \midrule
Login to Honeypots 1 attempt via SSH                                                                              & Attempt to breach security via online brute force                                                                                                                         & Medium-Low      & Auto or Human                  & Not observed.                                                                                                                      \\ \midrule
Login attempt in Web portal with any   username/password                                                          & Attempt to breach security via online brute force                                                                                                                         & Medium-High   & Auto or Human                     & Verified Correct.   Medium-High.       (Log of   usr/psw from PHP code)                                                                            \\ \midrule
Access attempt to disallowed entry in robot.txt                                                                   & Deliberate malicious attempt to expand reconnaissance   scope                                                                                                             & Medium-High   & Auto                    & Disallowed entry was not a factor in browsing the folder      Low.       (Access to /admin)                                                        \\ \midrule
Login attempt in Web portal with the honeytoken   credentials hidden as HTML comments                             & Attempt to breach security via online brute force   after examining the comments in the HTML. Indicative of human action.                                  & High           & Human                   & Verified Correct.       High.       (Log of   usr/psw from PHP code)                                                                               \\ \midrule
Login attempts in Web portal with variations of the   theme of the honeytoken credentials hidden as HTML comments & Attempt to breach security via online brute force   after examining the comments in the HTML and reasoning logical variations.   Indicative of human reasoning and action. & High-High     & Human                    & Verified Correct.       High-High.       (Log of   usr/psw from PHP code)                                                                          \\ \midrule
TCP SYN initiated from any honeypot        (breach of honeypot shell)                                             & Successful security breach. If honeypots have been   patched to latest version, implies possible exploitation of zero-day   vulnerability                                 & High-High (APT breach)   & Human         & Not Observed.                                                                                                                                      \\ \bottomrule
\end{tabular}
\end{center}
\caption{Honeytoken interactions and severity of the detected events}
\label{tab:result-table}
\vspace{-0.2cm}
\end{table*}

The results of the experiment reveal some interesting findings from the outputs of the monitoring functions and the interactions with the honeytokens. Some of the monitoring functions served to indicate an attack is being launched without revealing the threat level or severity of the intrusion, while others revealed clear intent and human interaction.

\begin{figure}[h]
    \centering
    \includegraphics[width=0.5\textwidth]{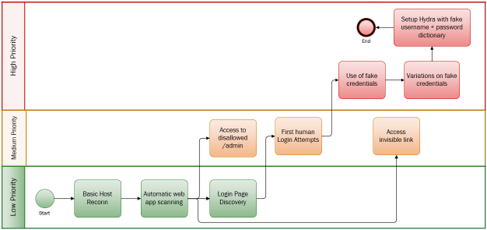}
    \caption{Attack State Diagram}
    \label{fig:attackstate}
\vspace{-0.1cm}
\end{figure}

In what follows all these cases will be examined. Figure~\ref{fig:attackstate}, shows a state diagram of the attack pattern indicating the indicator priority of each type of event detected.

\noindent \textbf{Low Priority Indicators:}
Any detected network or Web application interaction with the honeypot is indicative of reconnaissance activity by a potential attacker, but in some cases, it may be an opportunistic or an interaction by mistake by an insider with no malicious intent.

The detection of ICMP and TCP SYN packets directed at the honeypot both served to indicate this level of intrusion; they were one of the earliest signs of attack activity. There is nothing malicious about TCP SYN packets in company networks; but in this case, they indicate possible recon or TCP communication initiation to/from the honeypot; which is deemed unauthorised activity. Outgoing TCP SYN packets, would indicate a honeypot breach; a higher level of intrusion.

Signs of automatic Web app scanning and the discovery of the fake login Index.php page is also an early indication as can be seen from the profile of events.\\

\vspace{-0.1cm}
\noindent \textbf{Medium Priority Indicators:}
Initiation of a Web content scan implies that the host was identified as serving a Web page.

Access to the /admin folder which was included in the robot.txt file is a medium priority indicator, because this folder was detected directly by the initial Web scan process (Burpsuite) irrespective of its inclusion as a disallowed access within the robot.txt file.

Also, the sheer number of hits (a total of 56,114) indicated the use of an automated Web scanning tool
, as any normal user would not likely be accessing this folder so many times; however, the attack can still be opportunistic because it used a standard tool and required little human configuration or effort. 

Tests on the Web application carried out by the author using dirb revealed that the /admin folder was one of the first folders found by the tool, which then focused on the discovery of sub-folders within the /admin folder: This helps to explain why so many hits were detected.

A better implementation of this honeytoken might be the inclusion of less commonly named disallowed folders in robot.txt, one that is not part of dirb’s dictionaries or commonly used by Web developers, so that it would not have been scanned directly by common Web scanning tools using fuzzing techniques; but rather indicate an analysis of the contents of the robot.txt file.

Another medium priority indicator can be the access to the hidden link “/testlogin/index.php”; since it was not detected until stage 7 of the pentest; via the use of an automatic tool. The main advantage of this honeytoken is the fact that we can be almost certain that is was detected by automatic tool, because the link is invisible to the human eye. \\

\vspace{-0.1cm}
\noindent \textbf{High Priority Indicators:}
The honeytoken that resulted in the most effective high priority indicator was the fake credentials within the HTML page comments.
The use of these credentials indicated deliberate analysis of the contents of the HMTL source file and the introduction of variations in the credentials that were tested suggests the presence of a committed human actor, intent on breaching the security of the Web server.

The three login attempts with variations on the fake credentials were detected, including correcting a typo and adding the correct domain name to the fake user email; which could only be human attempts based on prior acquired knowledge.

This manual, low intensity, breach attempt was followed by an automated brute force attack towards the end of the pentest. This brute force attack is also a high priority indicator in that it takes human effort to configure and launch the attacking tool (in this case Hydra) pre-configured with the username previously obtained from the HTML comment honeytoken.

The brute force attack is easily detectable by the large number of login attempts (1406) vs the prior manual attempts or by other Web scanning tools such as Burpsuite (45).

The results of the experiment indicate that it is indeed possible to differentiate between automated attacks and those with human involvement, by means of the monitoring and honeytoken techniques deployed.

Table~\ref{tab:result-table} summarises the results of the observations and their implication in terms of automated vs human interactions, level of intent and severity. The experimental results verify that it was possible to segregate initial reconnaissance activities and automatic Web application scanning operations from human reasoned and initiated attacks within the security perimeter (RQ1). It was also possible to infer the level of intent of the attacker using the technique of planting  credentials that hinted at possible variations, such as completing a correct domain name (in this case adding a ``.co'' at the end), which in itself implies previous knowledge of the owner of the site being attacked (RQ2). This was also verified by the evidence of an automated brute force attack which was configured to use the fake username credentials and a dictionary of passwords. The fact that the attacker was led to try a further attack based on the honeytoken shows that the concept of breadcrumbs could be effective in detecting a structured attack. Indeed, this simple honeytoken technique (planted fake credentials) can be used, not only within HTML comments, but also within files, configured within domains, and as email accounts, as shown in previous work in the area of honey items.

These results demonstrate that it is possible to assign a priority or severity level to intrusion detection events by use of simple honeytokens and breadcrumbs. The concepts explored here could be improved and extended by: a) Using non-common folder names that cannot be fuzzed; b) Extending the use of breadcrumb honeytokens in production servers, giving clues that point to the honeypots or invite a login to servers; c) Allowing some of the fake credentials planted to be valid logins within honeypots, which can lead to the discovery of other honeytokens; thereby leading the attacker into a false trail of discovery, all the while being observed; and 
d) designing honeytokens that can be discovered by the documented MITRE ATT\&CK TTPs~\cite{MITRE2020}. 

\section{Conclusions}
\label{sec:conclusions}

The results of the experiment reveal an increment in the level of human reasoning and actions: While at the start, the human actor is launching automatic reconnaissance tools (Low priority), very soon this leads to the more detailed investigation of findings and even attempting a few manual attempts to login with any generic context-related credentials (Medium priority). The detection of these Low and Medium priority indicators starts to give a positive answer to RQ1. Finally, the discovery of the fake credential breadcrumb leads to manually testing them directly, creating plausible variations and setting up a brute force attack based on them, all of which demonstrate a high degree of human reasoning and a high level of intent (High priority), demonstrating both RQ1 and RQ2. Prior research focused on detecting automated attacks on login attempts due to their regularity or predictability and considering all other events as human driven, but the present work focuses directly on capturing those events which are very likely executed by a human showing that honeytokens are a very simple but effective technique in achieving this.  

If this type of deception is embedded within an organisation's intranet, it has the potential to detect the start of an APT or insider threat attack with very low false positives (as the attack intent and reality would be verified by the detection of ever higher priority indicators that demonstrate human actions) and low false negatives, as the honeytokens would be discoverable by any new unknown type of attack.
Future work can focus on extending the use of the most effective honeytoken techniques across production server and client hosts and allowing some of the fake credentials to initiate controlled sessions within honeypots to continue to observe the behaviour of the attacker. Future ambitious experiments that could reach more general conclusions could focus on setting up a completely fake company network on a public cloud, with fake user avatars that allow themselves to be hacked via phishing emails, in order to study the path of real APT actors in the wild from initial foothold onward.


\bibliographystyle{IEEEtran}
\bibliography{references_nourl}

\end{document}